# Quantum State Reduction and the Quantum Bayes Principle

Masanao Ozawa

*School of Informatics and Sciences, Nagoya University, Nagoya 464-01, Japan*

**Abstract**

This paper gives new foundations of quantum state reduction without appealing to the projection postulate for the probe measurement. For this purpose, the quantum Bayes principle is formulated as the most fundamental principle for determining the state of a quantum system, and the joint probability distribution for the outcomes of local successive measurements on a noninteracting entangled system is derived without assuming the projection postulate.

## 1 Introduction

In the discussion of new devices of measurement such as quantum nondemolition measurements and related proposals in the last two decades, the problem of the mathematical characterization of all the possible quantum measurements allowed in the standard formulation of quantum mechanics turned out to be of considerable potential importance in engineering and precision measurement experiments. When a new device of measurement is proposed, it is necessary, in general, to specify how the apparatus is prepared, how it interacts with the object, and how the outcome is obtained from it; these specifications will be called a model of measurement. On the other hand, a quantum measurement is specified from a statistical point of view by the outcome probability distribution and the state reduction, i.e., the state change from the state before measurement to the state after measurement conditional upon the outcome. If two measurements have the same outcome probability distribution and the same state reduction, they are statistically equivalent.

The conventional derivation of the state reduction from a given model of measurement is to compute the state of the object-apparatus composite system just after the measuring interaction and to apply the projection postulate to the subsequent measurement of the probe observable. This prescription is, however, not only controversial from the interpretational point of view but also even physically inconsistent. Some evidences of physical inconsistency can be pointed out as follows: (1) In some measuring apparatuses, the instrument for probe measurement such as a photon counter does not operate just as described by the projection postulate [IUO90]. (2) When the probe observable has continuous spectrum, the projection postulate to be applied cannot be formulated properly in the standard formulation of quantum mechanics [Oza84]. (3) State reduction should determine the state of the measured system at the instant just after the measuring interaction and just before the probe measurement, and hence the application of the projection postulate to the probe measurement is irrelevant to the state reduction [Oza89b, Oza95b].

A mathematically rigorous and physically consistent derivation of the state reduction from any model of measurement without applying the projection postulate to the probe measurement has been established in [Oza83, Oza84, Oza85a]. Based on this derivation, the problem of the mathematical characterization of all the possible quantum measurements was solved as follows [Oza83, Oza84]: A measurement, with the outcome distribution $P(dx|\rho)$ and the state reduction $\rho \mapsto \rho_x$, is realizable in the standard formulation of quantum mechanics if and only if its statistics are representable by a normalized completely positive (CP) map valued measure $\mathbf{X}$ in such a way that $\mathbf{X}(dx)\rho = \rho_x P(dx|\rho)$, where the CP maps $\mathbf{X}(\Delta)$ are defined on the space of trace class operators for all Borel subsets $\Delta$ of the space of possible outcomes. The statistical equivalence classes of measurements are thus characterized as the normalized CP map valued measures.

In this paper, I will discuss further the foundations of quantum state reduction. The quantum Bayes principle will be formulated as the most fundamental principle for determining the state of a quantum system. The joint probability distribution will be also derived for the outcomes of local successive measurements on a noninteracting entangled system without assuming the projection postulate. This joint probability distribution and the quantum Bayes principle will naturally lead to the state reduction for an arbitrary model of measurement.

For simplicity we will be confined to measurements of *discrete observables*, but it will be easy for the reader to generalize the argument to continuous observables and to join the argument to the general theory developed in such papers as [Oza84, Oza85a, Oza85b, Oza86, Oza89a, Oza91, Oza93, Oza95a].

## 2  Quantum Bayes Principle

Let $X, Y$ be two (discrete) random variables. Suppose that we know the joint probability distribution $\Pr\{X = x, Y = y\}$. Then, the prior distribution of $X$ is defined as the marginal distribution of $X$, i.e.,

$$\Pr\{X = x\} = \sum_y \Pr\{X = x, Y = y\}. \tag{1}$$

If one measures $Y$, the *information* $Y = y$ changes the probability distribution of $X$ for any outcome $y$. The posterior distribution of $X$ is defined as the conditional probability distribution of $X$ given $Y = y$, i.e.,

$$\Pr\{X = x | Y = y\} = \frac{\Pr\{X = x, Y = y\}}{\sum_x \Pr\{X = x, Y = y\}}. \tag{2}$$

This method of changing the probability distribution from the prior distribution to the posterior distribution is called as the Bayes principle. The Bayes principle is one of the most fundamental principle in the statistical inference.

In quantum mechanics, the notion of probability distribution is related to the notion of state. Roughly speaking, "the state of the system" is equivalent to "the probability distributions of all the observables of the system". Thus the Bayes principle yields the state change of a quantum system if the probability distributions of all the observables of the system has changed by the Bayes principle. We formulate this principle of state changes as follows:

**The Quantum Bayes Principle:** If an information changes the probability distributions of *all* the observables of a quantum system according to the Bayes principle, then the information changes the *state* of the system according to the change of the probability distributions.

## 3 Quantum Rules

By the quantum rules we mean the following three basic principles in nonrelativistic quantum mechanics.

1. *Schrödinger Equation*: The time evolution of the system is given by

$$\psi \mapsto e^{-iH\tau/\hbar}\psi.$$

where $H$ is the Hamiltonian of the system.

2. *Statistical Formula*: The probability distribution of the outcome of the measurement of an observable $A$ in the state $\psi$ is given by

$$\Pr\{A = a\} = \|E^A(a)\psi\|^2$$

where $E^A(a)$ denotes the projection operator with the range $\{\psi \in \mathcal{H}|\ A\psi = a\psi\}$—if $a$ is an eigenvalue of $A$, it is the spectral projection corresponding to $a$; otherwise $E^A(a) = 0$.

3. *Projection Postulate* [Lud51]: The state change caused by the measurement of an observable $A$ with the outcome $a$ is given by

$$\psi \mapsto \frac{E^A(a)\psi}{\|E^A(a)\psi\|^2}.$$

From the above quantum rules, we can deduce

4. *Joint Probability Distributions of Successive Measurements* [Wig63]: If any sequence of observables $A_1, \ldots, A_n$ in a system originally in the state $\psi$ are measured at the times $0 \leq t_1 < \cdots < t_n$, then the joint probability distribution of the outcomes is given by

$$\Pr\{A_1(t_1) = a_1, \ldots, A_n(t_n) = a_n\} = \|E^{A_n}(a_n)e^{-iH(t_n-t_{n-1})/\hbar}\cdots E^{A_1}(a_1)e^{-iHt_1/\hbar}\psi\|^2.$$

By rule 4 with $0 = t_1 < \cdots < t_n \approx 0$, we obtain

5. *Simultaneous Measurability of Commuting Observables*: Mutually commuting observables $A_1, \ldots, A_n$ are simultaneously measurable. The joint probability distribution of the outcomes in the state $\psi$ is given by

$$\Pr\{A_1 = a_1, \ldots, A_n = a_n\} = \|E^{A_1}(a_1) \cdots E^{A_n}(a_n)\psi\|^2.$$

## 4 Difficulties in the Projection Postulate

As shown above, the projection postulate plays one of the most fundamental roles in foundations of quantum mechanics. The following difficulties in this postulate, however, have been pointed out among others:

1. There are measurements of an observable $A$ which does not satisfy the projection postulate.

2. If $A$ has a continuous spectrum, we have no projection postulate for the measurement of $A$ [Oza84, Oza85a].

In order to illustrate the measurement of an observable which does not satisfy the projection postulate, consider a model of measurement in which the object interacts with the apparatus for a finite time interval. Let $A = \sum_n a_n |\phi_n\rangle\langle\phi_n|$ be the observable to be measured and $B = \sum_m b_m |\xi_m\rangle\langle\xi_m|$ the probe observable in the apparatus. Let $\xi$ be the apparatus initial state and $U$ the unitary evolution of the object-apparatus composite system under measuring interaction. The measuring interaction transduces the observable $A$ to the probe observable $B$ and the outcome of the measurement is obtained by amplifying $B$ after the interaction, in the subsequent stage of the apparatus, to the directly sensible extent. Then we have:

1. The measurement *with* the projection postulate is described by

$$U : \phi_n \otimes \xi \mapsto \phi_n \otimes \xi_n.$$

2. The measurement *without* the projection postulate is described by

$$U : \phi_n \otimes \xi \mapsto \phi'_n \otimes \xi_n$$

where $\{\phi'_n\}$ is an *arbitrary* family of states.

By linearity, the unitary $U$ in 2 satisfies

$$U : \left(\sum_n c_n \phi_n\right) \otimes \xi \mapsto \sum_n c_n \phi'_n \otimes \xi_n.$$

Hence, the outcome probability distribution of the measurement with process 2, which is obtained by the probability distribution of the probe observable, satisfies, in fact, the statistical formula for the observable $A$.

A typical example of measurement without the projection postulate is the photon counting. In this case, the measured observable is the number operator $A = \hat{n} = \sum_n n|n\rangle\langle n|$ and the time evolution of the composite system can be described by

$$U : |n\rangle \otimes \xi \mapsto |0\rangle \otimes \xi_n,$$

in the idealized model—for a more detailed model, see [IUO90].

## 5 Local Measurement Theorem

Let $\mathbf{S}_1$ be a quantum system with the free Hamiltonian $H_1$ and the Hilbert space $\mathcal{H}_1$ and let $\mathbf{S}_2$ a system with the free Hamiltonian $H_2$ and the Hilbert space $\mathcal{H}_2$. Let $A$ be an observable of the system $\mathbf{S}_1$ and $B$ an observable of the system $\mathbf{S}_2$. Suppose that the composite system $\mathbf{S}_{12} = \mathbf{S}_1 + \mathbf{S}_2$ is initially in the state (density operator) $\rho(0) = \rho$. Suppose that at the time $t_1$ the observable $A$ is measured by an apparatus $\mathbf{A}$, at the time $t_2$ ($0 < t_1 < t_2$) the observable $B$ is measured by any apparatus measuring $B$, and that there is no interaction between $\mathbf{S}_1$ and $\mathbf{S}_2$—namely, the system $\mathbf{S}_{12}$ is a noninteracting entangled system. Denote the joint probability distribution of the outcomes of the $A$-measurement and the $B$-measurement by

$$\Pr\{A(t_1) = a, B(t_2) = b\|\rho\}.$$

According to rule 4, if the $A$-measurement satisfies the projection postulate, the joint probability distribution is given by

$$\Pr\{A(t_1) = a, B(t_2) = b\|\rho\} = \operatorname{Tr}\left[\left(e^{iH_1t_1/\hbar}E^A(a)e^{-iH_1t_1/\hbar} \otimes e^{iH_2t_2/\hbar}E^B(b)e^{-iH_2t_2/\hbar}\right)\rho\right]. \tag{3}$$

In what follows we shall derive the above formula *without* assuming the projection postulate.

We note that this joint probability distribution should be affine in the state $\rho$. To show this, suppose that the state $\rho$ is the mixture of states $\rho_1$ and $\rho_2$

$$\rho = \alpha\rho_1 + (1-\alpha)\rho_2 \tag{4}$$

where $0 < \alpha < 1$. This means that the measured system $\mathbf{S}_{12}$ is a random sample from the ensemble with the density operator $\rho_1$ with probability $\alpha$ and from the ensemble with the density operator $\rho_2$ with probability $1 - \alpha$. Hence we have

$$\begin{aligned}&\Pr\{A(t_1) = a, B(t_2) = b\|\rho\} \\ &= \alpha \Pr\{A(t_1) = a, B(t_2) = b\|\rho_1\} + (1-\alpha)\Pr\{A(t_1) = a, B(t_2) = b\|\rho_2\}.\end{aligned} \tag{5}$$

Next, we introduce an important condition for the measuring apparatus which leads to the desired formula (3). We say that the measuring apparatus $\mathbf{A}$ is *local* at the system $\mathbf{S}_1$ if the measuring interaction occurs only in the apparatus and the system $\mathbf{S}_1$, or precisely, if the operator representing the measuring interaction commutes with every observable of $\mathbf{S}_2$. If this is the case, the total Hamiltonian of the composite system $\mathbf{A} + \mathbf{S}_1 + \mathbf{S}_2$ during the measuring interaction is represented by

$$H_{tot} = H_{\mathbf{A}} \otimes 1 \otimes 1 + 1 \otimes H_1 \otimes 1 + 1 \otimes 1 \otimes H_2 + KH_{int} \otimes 1 \tag{6}$$

where $H_{\mathbf{A}}$ is the free Hamiltonian of the apparatus, $H_{int}$ is the operator on $\mathcal{H}_{\mathbf{A}} \otimes \mathcal{H}_1$ representing the measuring interaction, where $\mathcal{H}_{\mathbf{A}}$ is the Hilbert space of the apparatus, and $K$ is the coupling constant. Then, we can show that there is a unitary operator $U$ on the Hilbert space $\mathcal{H}_{\mathbf{A}} \otimes \mathcal{H}_1$ such that the state of the system $\mathbf{S}_{12}$ at the time $t + \Delta t$, where $t$ is the time of measurement and $\Delta t$ is the duration of measuring interaction, is obtained by

$$\rho(t + \Delta t) = \text{Tr}_{\mathbf{A}} \left[ (U \otimes e^{-iH_2\Delta t/\hbar}) (\sigma \otimes \rho(t)) (U^\dagger \otimes e^{iH_2\Delta t/\hbar}) \right]. \tag{7}$$

where $\sigma$ is the prepared state of the apparatus at the time of measurement. In fact, $U$ is given by

$$U = e^{-i(1 \otimes H_1 + KH_{int})\Delta t/\hbar}.$$

Now, we shall prove the following theorem on the joint probability distribution of the outcomes of the $A$-measurement and the $B$-measurement.

**Theorem 5.1 (Local Measurement Theorem)** *If the measuring apparatus $\mathbf{A}$ measuring $A$ is local at the system $\mathbf{S}_1$, the joint probability distribution of the outcomes of the $A$-measurement and the $B$-measurement is given by*

$$\Pr\{A(t_1) = a, B(t_2) = b \| \rho\} = \text{Tr}\left[ \left( e^{iH_1 t_1/\hbar} E^A(a) e^{-iH_1 t_1/\hbar} \otimes e^{iH_2 t_2/\hbar} E^B(b) e^{-iH_2 t_2/\hbar} \right) \rho \right]. \tag{8}$$

*Proof.* For any real numbers $a, b$ and any density operator $\rho$ on $\mathcal{H}_1 \otimes \mathcal{H}_2$, let

$$P(a, b, \rho) = \Pr\{A(t_1) = a, B(t_2) = b \| \rho\}.$$

By (5), the function $\rho \mapsto P(a, b, \rho)$ is a positive affine function on the space of density operators on $\mathcal{H}_1 \otimes \mathcal{H}_2$. Since the convex set of density operators is a base of the base norm space of trace class operators, this affine function is extended uniquely to a positive linear functional on the space of trace class operators. By the Schatten-von Neumann duality theorem, the space of bounded operators is the dual space of the space of trace class operators, and hence there is a positive operator $F(a, b)$ on $\mathcal{H}_1 \otimes \mathcal{H}_2$ such that

$$P(a, b, \rho) = \text{Tr}[F(a, b)\rho].$$

For any $\rho$ we have

$$\text{Tr}\left[\sum_b F(a,b)\rho\right] = \Pr\{A(t_1) = a\|\rho\} = \text{Tr}\left[\left(e^{iH_1t_1/\hbar}E^A(a)e^{-iH_1t_1/\hbar} \otimes 1\right)\rho\right],$$

and hence we have

$$\sum_b F(a,b) = e^{iH_1t_1/\hbar}E^A(a)e^{-iH_1t_1/\hbar} \otimes 1. \tag{9}$$

By the locality condition (7), for any $\rho$ we have

$$\text{Tr}\left[\sum_a F(a,b)\rho\right]$$
$$= \Pr\{B(t_2) = b\|\rho\}$$
$$= \text{Tr}\left[\left(1 \otimes E^B(b)\right)\rho(t_2)\right]$$
$$= \text{Tr}\left[\left(1 \otimes e^{iH_2\tau/\hbar}E^B(b)e^{-iH_2\tau/\hbar}\right)\text{Tr}_{\mathbf{A}}\left[\left(U \otimes e^{-iH_2\Delta t/\hbar}\right)(\sigma \otimes \rho(t_1))\left(U^\dagger \otimes e^{iH_2\Delta t/\hbar}\right)\right]\right]$$
$$= \text{Tr}\left[\left(1 \otimes e^{iH_2t_2/\hbar}E^B(b)e^{-iH_2t_2/\hbar}\right)\rho\right]$$

where $\tau = t_2 - t_1 - \Delta t$ and hence

$$\sum_a F(a,b) = 1 \otimes e^{iH_2t_2/\hbar}E^B(b)e^{-iH_2t_2/\hbar}. \tag{10}$$

Since every positive operator valued measure on a product space with projection valued marginal measures is the product of its marginal measures [Dav76, page 39], by (9) and (10) we have

$$F(a,b) = e^{iH_1t_1/\hbar}E^A(a)e^{-iH_1t_1/\hbar} \otimes e^{iH_2t_2/\hbar}E^B(b)e^{-iH_2t_2/\hbar}.$$

Therefore, (8) follows. □

## 6 Quantum state reduction

Consider a model of measurement on a system **S** at the time $t$. Let **A** be the apparatus with the probe observable $A$. The measurement is carried out by the interaction between **S** and **A** from the time $t$ to the time $t + \Delta t$. The object **S** is free from the apparatus **A** after the time $t + \Delta t$. Suppose that at the time $t$ the object **S** is in the state $\rho(t)$ and that the apparatus **A** is prepared in the state $\sigma$. Let $U$ be the unitary operator representing the time evolution of the object-probe composite system **A** + **S** from the time $t$ to $t + \Delta t$. Then the system **A** + **S** is in the state $U(\sigma \otimes \rho(t))U^\dagger$ at the time $t + \Delta t$. The outcome of this measurement is obtained by the measurement, local at the system **A**, of the probe observable $A$ at the time $t + \Delta$. Hence, the probability distribution of the outcome **a** of this measurement is given by

$$\Pr\{\mathbf{a} = a\} = \Pr\{A(t + \Delta t) = a\} = \text{Tr}\left[\left(E^A(a) \otimes 1\right)U(\sigma \otimes \rho(t))U^\dagger\right]. \tag{11}$$

In order to determine the state reduction caused by this measurement, suppose that the observer were to measure an arbitrary observable $B$ of the object **S** at the time $t + \Delta t + \tau$ with $\tau \geq 0$. Then the joint probability distribution of the outcome **a** and the outcome $B(t + \Delta t + \tau)$ of the $B$-measurement at $t + \Delta t + \tau$ is identical with the joint probability distribution of the outcomes of the $A$-measurement at $t + \Delta t$ and the $B$-measurement at $t + \Delta t + \tau$, i.e.,

$$\Pr\{\mathbf{a} = a, B(t + \Delta t + \tau) = b\} = \Pr\{A(t + \Delta t) = a, B(t + \Delta t + \tau) = b\}. \tag{12}$$

By the local measurement theorem, we have

$$\Pr\{A(t + \Delta t) = a, B(t + \Delta t + \tau) = b\}$$
$$= \text{Tr}\left[\left(E^A(a) \otimes e^{iH\tau/\hbar} E^B(b) e^{-iH\tau/\hbar}\right) U(\sigma \otimes \rho(t)) U^\dagger\right]. \tag{13}$$

Thus, the prior probability distribution of $B(t + \Delta t + \tau)$ is the marginal distribution of $B(t + \Delta t + \tau)$, i.e.,

$$\begin{aligned}
\Pr\{B(t + \Delta t + \tau) = b\} &= \sum_a \Pr\{A(t + \Delta t) = a, B(t + \Delta t + \tau) = b\} \\
&= \text{Tr}\left[\left(1 \otimes e^{iH\tau/\hbar} E^B(b) e^{-iH\tau/\hbar}\right) U(\sigma \otimes \rho(t)) U^\dagger\right] \\
&= \text{Tr}\left[e^{iH\tau/\hbar} E^B(b) e^{-iH\tau/\hbar} \text{Tr}_\mathbf{A}\left[U(\sigma \otimes \rho(t)) U^\dagger\right]\right] \tag{14}
\end{aligned}$$

where $\text{Tr}_\mathbf{A}$ is the partial trace over the Hilbert space of the apparatus. Thus we can define the prior state of the system **S** at the time $t + \Delta t$ by

$$\rho(t + \Delta t) = \text{Tr}_\mathbf{A}\left[U(\sigma \otimes \rho(t)) U^\dagger\right], \tag{15}$$

which describes the prior probability distributions of all observable $B$ of the system **S** after the time $t + \Delta t$. Since this state change $\rho(t) \mapsto \rho(t + \Delta t)$ does not depend on the outcome of the measurement, this process is called the *nonselective measurement*.

If one reads out the outcome **a**, or $A(t + \Delta t)$, of this measurement, the information $\mathbf{a} = a$ changes the probability distribution of $B(t + \Delta t + \tau)$ for any outcome $a$ from the prior distribution to the posterior distribution according to the Bayes principle. The posterior distribution of $B(t+\Delta t+\tau)$ is defined as the conditional probability distribution of $B(t + \Delta t + \tau)$ given $\mathbf{a} = a$, i.e.,

$$\begin{aligned}
&\Pr\{B(t + \Delta t + \tau) = b | \mathbf{a} = a\} \\
&= \frac{\Pr\{\mathbf{a} = a, B(t + \Delta t + \tau) = b\}}{\Pr\{\mathbf{a} = a\}} \\
&= \frac{\text{Tr}\left[\left(E^A(a) \otimes e^{iH\tau/\hbar} E^B(b) e^{-iH\tau/\hbar}\right) U(\sigma \otimes \rho(t)) U^\dagger\right]}{\text{Tr}\left[(E^A(a) \otimes 1) U(\sigma \otimes \rho(t)) U^\dagger\right]} \\
&= \frac{\text{Tr}\left[e^{iH\tau/\hbar} E^B(b) e^{-iH\tau/\hbar} \text{Tr}_\mathbf{A}[(E^A(a) \otimes 1) U(\sigma \otimes \rho(t)) U^\dagger]\right]}{\text{Tr}\left[(E^A(a) \otimes 1) U(\sigma \otimes \rho(t)) U^\dagger\right]}. \tag{16}
\end{aligned}$$

Thus, letting

$$\rho(t+\Delta t|\mathbf{a}=a) = \frac{\mathrm{Tr}_{\mathbf{A}}\left[\left(E^A(a)\otimes 1\right)U\left(\sigma\otimes\rho(t)\right)U^\dagger\right]}{\mathrm{Tr}\left[(E^A(a)\otimes 1)U\left(\sigma\otimes\rho(t)\right)U^\dagger\right]}, \quad (17)$$

we have

$$\Pr\{B(t+\Delta t+\tau)=b|\mathbf{a}=a\} = \mathrm{Tr}[e^{iH\tau/\hbar}E^B(b)e^{-iH\tau/\hbar}\rho(t+\Delta t|\mathbf{a}=a)]. \quad (18)$$

This shows that the posterior distribution of the outcome of the measurement of *any* observable $B$ of the object **S** after the time $t+\Delta t$ is described by the state $\rho(t+\Delta t|\mathbf{a}=a)$. Therefore, we can conclude that the information $\mathbf{a}=a$ changes the state of the system **S** at the time $t+\Delta t$ from the *prior state* $\rho(t+\Delta t)$ to the *posterior state* $\rho(t+\Delta t|\mathbf{a}=a)$ according to the quantum Bayes principle.

The state reduction $\rho(t) \mapsto \rho(t+\Delta t|\mathbf{a}=a)$ is thus obtained as the composition of the state change $\rho(t) \mapsto \rho(t+\Delta t)$ by the measuring interaction and the state change $\rho(t+\Delta t) \mapsto \rho(t+\Delta t|\mathbf{a}=a)$ by the information on the outcome of the measurement.

## 7 Conclusion

We have formulated the quantum Bayes principle and proved the local measurement theorem. These theoretical foundations lead to the following new derivation of state reduction. From the time $t$ of measurement to the time $t+\Delta t$ just after measurement, the object **S** interacts with the apparatus **A**. Thus the state of the object changes dynamically

$$\rho(t) \mapsto \rho(t+\Delta t) = \mathrm{Tr}_{\mathbf{A}}\left[U\left(\sigma\otimes\rho(t)\right)U^\dagger\right]. \quad (19)$$

This process is the nonselective measurement, which does not depends on the outcome of the measurement. The state reduction is the state change of the object from the time $t$ to the time $t+\Delta t$ depending upon the outcome $\mathbf{a}$. According to the quantum Bayes principle, the information $\mathbf{a}=a$ changes the state of the object at the time $t+\Delta t$ from the prior state $\rho(t+\Delta t)$ to the posterior state $\rho(t+\Delta t|\mathbf{a}=a)$, i.e.,

$$\rho(t+\Delta t) \mapsto \rho(t+\Delta t|\mathbf{a}=a) = \frac{\mathrm{Tr}_{\mathbf{A}}[(E^A(a)\otimes 1)U(\sigma\otimes\rho(t))U^\dagger]}{\mathrm{Tr}[(E^A(a)\otimes 1)U(\sigma\otimes\rho(t))U^\dagger]}. \quad (20)$$

This change of state includes no dynamical element. Thus the state reduction is obtained as the composition of the dynamical change $\rho(t) \mapsto \rho(t+\Delta t)$ and the *informatical* change $\rho(t+\Delta t) \mapsto \rho(t+\Delta t|\mathbf{a}=a)$.

The above derivation does not assume the projection postulate for the probe measurement. Formula (20) shows that the state after measurement conditional upon the outcome of the measurement does not depend on whether the probe measurement satisfies the projection postulate or not. Thus, formula (20) applies to any measurements whose probe measurement may not satisfy the projection postulate such as photon counting.